\documentclass[conference]{IEEEtran}
\usepackage{fancyhdr}
\IEEEoverridecommandlockouts
\usepackage{cite}
\usepackage{amsmath,amssymb,amsfonts}
\usepackage{graphicx}
\usepackage{textcomp}
\usepackage{xcolor}
\usepackage{tikz}
\usepackage{algorithm}
\usepackage{algpseudocode}
\usepackage{cleveref}
\usepackage{booktabs}
\usepackage{enumitem}
\usepackage{multirow}
\usepackage{xspace}
\usepackage{subcaption}
\def\BibTeX{{\rm B\kern-.05em{\sc i\kern-.025em b}\kern-.08em
    T\kern-.1667em\lower.7ex\hbox{E}\kern-.125emX}}

\newcommand{\sysname}{BaGuaLu-2\xspace}
\newcommand{\rev}[1]{#1}

\begin{document}

\title{\sysname{}: Taming Dynamic Memory Fluctuation for Efficient MoE Model Training \\over 10,000 GPUs
}


\author{
\IEEEauthorblockN{Mingshu Zhai}
\IEEEauthorblockA{\textit{Tsinghua University} \\
Beijing, China \\
zhaims22@mails.tsinghua.edu.cn}
\and
\IEEEauthorblockN{Minkun Guo}
\IEEEauthorblockA{\textit{Tsinghua University} \\
\textit{Qingcheng.AI} \\
Beijing, China \\
gmk21@mails.tsinghua.edu.cn}
\and
\IEEEauthorblockN{Hongrui Lian}
\IEEEauthorblockA{\textit{Tsinghua University} \\
\textit{Qingcheng.AI} \\
Beijing, China \\
lianhr24@mails.tsinghua.edu.cn}
\and
\IEEEauthorblockN{Shuangyu Li}
\IEEEauthorblockA{\textit{Tsinghua University} \\
Beijing, China \\
lishuang22@mails.tsinghua.edu.cn}
\and
\IEEEauthorblockN{Qi Zhang}
\IEEEauthorblockA{\textit{Tsinghua University} \\
Beijing, China \\
qi\_zhang@tsinghua.edu.cn}
\and
\IEEEauthorblockN{Chen Hu}
\IEEEauthorblockA{\textit{Unaffiliated} \\
Beijing, China \\
hclearner2@gmail.com}
\and
\IEEEauthorblockN{Xiawei Du}
\IEEEauthorblockA{\textit{Tsinghua University} \\
Beijing, China \\
duxw25@mails.tsinghua.edu.cn}
\and
\IEEEauthorblockN{Yuyang Jin\IEEEauthorrefmark{1}\thanks{*Corresponding authors}}
\IEEEauthorblockA{\textit{Tsinghua University} \\
Beijing, China \\
jinyuyang@tsinghua.edu.cn}
\and
\IEEEauthorblockN{Zan Zong\IEEEauthorrefmark{1}}
\IEEEauthorblockA{\textit{University of Science and Technology Beijing} \\
Beijing, China \\
zongzan@ustb.edu.cn}
\and
\IEEEauthorblockN{Jidong Zhai\IEEEauthorrefmark{1}}
\IEEEauthorblockA{\textit{Tsinghua University} \\
Beijing, China \\
zhaijidong@tsinghua.edu.cn}
}

\maketitle

\thispagestyle{fancy}
\lhead{}\rhead{}\chead{}
\lfoot{}
\rfoot{}\cfoot{}
\renewcommand{\headrulewidth}{0pt}
\renewcommand{\footrulewidth}{0pt}

\begin{abstract}

Fine-grained Mixture-of-Experts (MoE) is the de facto architecture for scaling large language models. However, dynamic expert load imbalance causes severe memory fluctuations that degrade training efficiency. Existing systems rely on static memory provisioning: they either over-provision, leaving GPU memory idle during normal execution, or under-provision, inviting out-of-memory crashes and costly manual rollbacks when load spikes occur. 
To address these challenges, we propose \sysname{}, a large-scale MoE training system that eliminates this dilemma through flexible provisioning. By transforming memory allocation from a static, one-time tuning choice into a continuous and fine-grained control. \sysname{} grants exact activation memory on demand for each training batch. Furthermore, it offsets any dynamic management overhead through system optimizations. \sysname{} demonstrates exceptional scalability, training a model with 1040 billion parameters across 12,288 GPUs while sustaining over 2 EFlops at 35\% MFU. Compared with state-of-the-art systems, \sysname{} delivers 1.14$\times$ and 1.13$\times$  average throughput improvements over Megatron-Kuai and AdaPipe, respectively.



\end{abstract}

\section{Introduction}
\label{sec:introduction}

Large language models (LLMs) have demonstrated remarkable capabilities across domains,
driving an ever-growing demand for larger and more capable models.
To meet this demand, fine-grained Mixture-of-Experts (MoE) has emerged as the dominant architecture for scaling
LLMs~\cite{deepseekv3,deepseekr1,qwen3,kimik2,glm4.5,gptoss,glm5},
as it increases model capacity without proportionally increasing the computation per token. Accelerating MoE training process has become a major focus in current systems research.

A fundamental characteristic of the MoE architecture is its dynamic sparse routing mechanism, which inherently introduces expert load imbalance. 
We observe this phenomenon consistently when scaling training to clusters comprising tens of thousands of GPUs. 
In approximately $2.9\%$ of training iterations, imbalanced load exhibits maximum $2.6\times$ spikes.
The imbalanced load leads to dynamic memory fluctuation,
where training jobs have the risk to exceed physical hardware capacity and triggering Out-of-Memory (OOM) errors, which halt the training process.
In response to the huge memory requirement of large-scale model training,
existing systems optimize GPU memory usage from multiple angles.
For weight and optimizer-states,
model parallelism~\cite{megatron} and ZeRO~\cite{zero} effectively shards these memory usage across devices.
For activation memory, systems rely on recomputation~\cite{adapipe} and offloading~\cite{megatron-kuai}
to reduce peak consumption.
However, these static approaches fail to resolve dynamic memory fluctuation introduced by load imbalance.

\begin{figure}[t]
    \centering
    \begin{subfigure}{\linewidth}
        \centering
        \includegraphics[width=\linewidth]{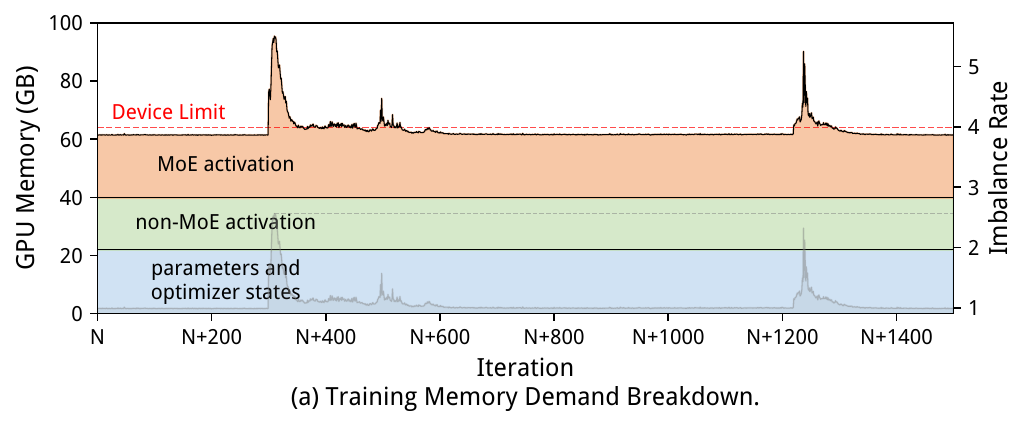}
    \end{subfigure}    
    \begin{subfigure}{\linewidth}
        \centering
        \includegraphics[width=\linewidth]{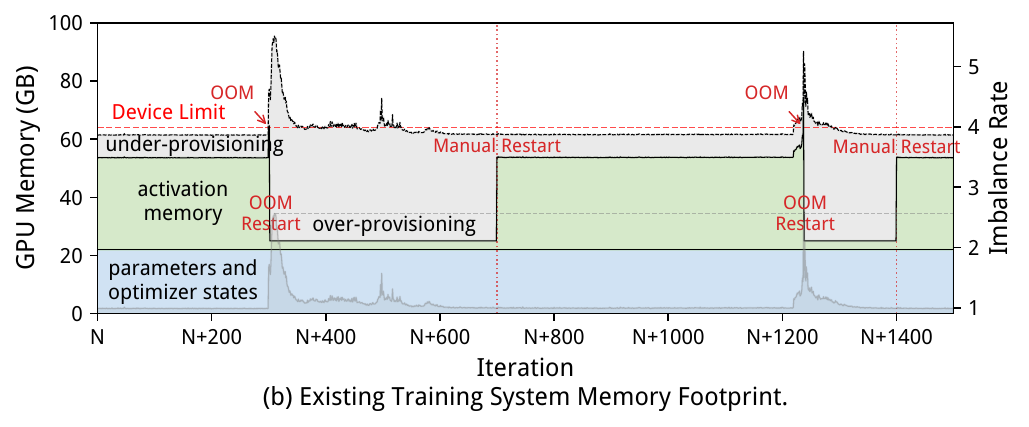}
    \end{subfigure}
    \vspace{-1em}
    \begin{subfigure}{\linewidth}
        \centering
        \includegraphics[width=\linewidth]{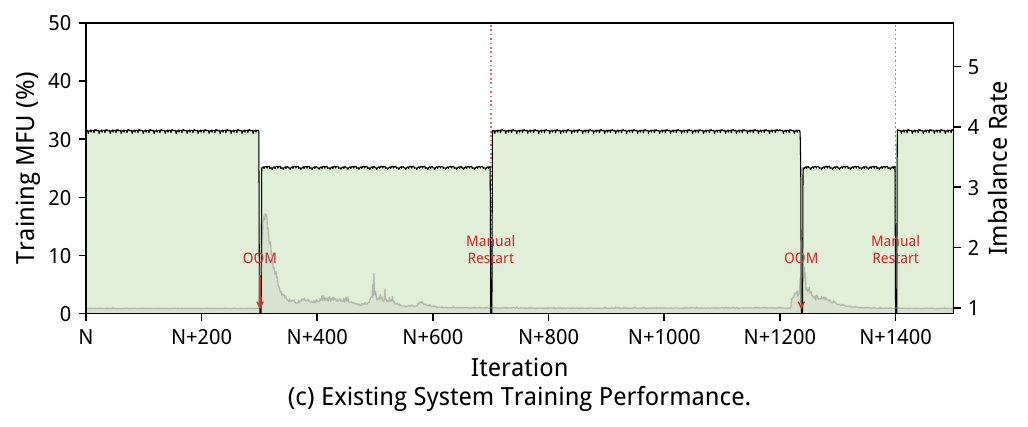}
    \end{subfigure}
    \caption{Dynamic memory fluctuation of MoE model training.}
    \vspace{-2em}
    \label{fig:main}
\end{figure}

Existing training systems are trapped in a rigid dilemma:
as shown in Figure~\ref{fig:main}(b),
if the system \emph{over-provisions} memory to avoid OOM errors,
GPU memory is wasted during most of the iterations, resulting in persistently low Model FLOPs Utilization~(MFU).
Conversely, if the system \emph{under-provisions} memory to maximize MFU,
it cannot dynamically respond to memory spikes at runtime.
Upon encountering a spike, the system crashes and requires manual intervention.
Restart and set the system in a conservative, over-provisioning memory configuration.
Once the spike passes and the operator wishes to restore the high-throughput
under-provisioned configuration, yet another manual restart is required.
Every such restart incurs significant performance overhead.

We identify activation memory as the root cause of both the memory bottleneck and the fluctuation
at scale.
Figure~\ref{fig:main}(a) shows the per-GPU memory demand breakdown in large-scale training.
Weight and optimizer-state memory can be distributed across GPUs and remains stable across iterations.
Activation memory, in contrast, cannot be effectively distributed---larger clusters require
proportionally more input data to maintain compute density---and
fluctuates unpredictably due to expert load imbalance caused by MoE's dynamic sparse routing.
Thus, dynamically managing activation memory is the key to addressing
the efficiency and stability challenges of large-scale MoE training.
However, existing systems' activation memory management suffers from two fundamental deficiencies.

\noindent\textbf{Static Memory Provisioning.}
Existing systems treat activation provisioning as a coarse-grained, fixed policy:
the policy is determined before training and cannot be adjusted at runtime.
For example, 
although AdaPipe~\cite{adapipe} supports adaptive recomputation across pipeline stages, 
each pipeline stage remains a fixed strategy throughout training.
This prevents the system from responding to per-batch memory fluctuation.

\noindent\textbf{Large Memory Mitigation Overhead.}
The extra computation or communication introduced by
activation mitigation is directly exposed on the critical path
and cannot be overlapped.
For example, the activation offloading in Megatron-Kuai~\cite{megatron-kuai}
achieves only limited overlap at inter-layer boundaries.
Activation mitigation therefore becomes a pure performance penalty,
and systems tend to minimize its use,
further limiting the scope for memory optimization.

\begin{figure}[t]
    \centering
    \includegraphics[width=\linewidth]{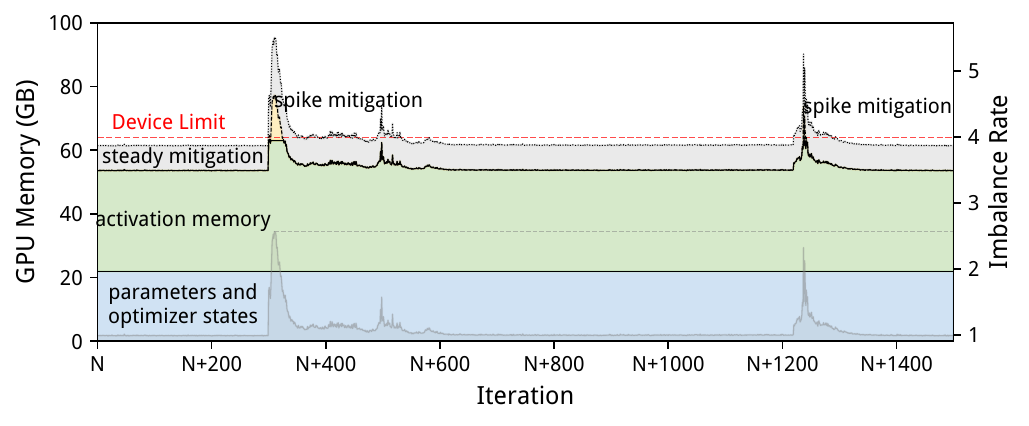}
    \vspace{-2em}
    \caption{Training Memory Footprint with \textit{flexible-provisioning}.}
    \vspace{-1.5em}
    \label{fig:bgl-mem}
\end{figure}

Constrained by these deficiencies, existing systems can only make a static binary choice
between over-provision and under-provision.
They are unable to dynamically allocate exactly the resources needed based on
the actual memory fluctuation of each training iteration.
We propose \emph{flexible activation provisioning},
which transform activation memory provisioning
from a \emph{static choice} into \emph{continuous control},
thereby achieving fully-utilized GPU memory.
As shown in \Cref{fig:bgl-mem},
this enables the system to tame dynamic memory fluctuation in both regimes:
\emph{(i)}~when MoE load imbalance is moderate,
the system applies steady activation mitigation to enable high-throughput parallelism, consuming almost all available memory;
\emph{(ii)}~when load imbalance is severe,
the system dynamically intensifies activation mitigation to prevent OOM,
avoiding the far greater cost of a cold restart.
We design three techniques to realize flexible-provisioning at training runtime.

\noindent\textbf{Activation Memory Modeling (\S\ref{sec:model}).}
We design an activation memory cost model,
which can accurately predict activation memory usage and mitigation time consumption.
It guides the following runtime memory provisioning strategies.

\noindent\textbf{Adaptive Activation Provisioning (\S\ref{subsec:provisioning}).}
Considering pipeline schedule phase and current MoE imbalance load,
we apply flexible memory budget for each micro batch.
Within a batch,
operators can use different mitigation strategies,
giving the system continuous control over memory consumption.


\noindent\textbf{Fused Dual-Batch Execution (\S\ref{subsec:overlap}).}
We propose interleaved execution for adjacent micro batch
to avoid directly exposing activation mitigation overhead on the critical path.
Using a light-weight runtime solver,
mitigation operations are inserted into hardware bubble 
so that the execution overhead is hidden.


These techniques are implemented in \sysname{},
a large-scale MoE training system.
We scale up to 12,288 GPUs for training 1040 billion parameters MoE model. 
\sysname{} achieves throughput of over 2~EFlops and an MFU of 35\%.
Compared with state-of-the-art training systems Megatron-Kuai~\cite{megatron-kuai} and AdaPipe~\cite{adapipe},
\sysname{} improves both training stability and peak performance,
with average throughput gains of $1.14\times$ and $1.13\times$, respectively.



\section{Background and Motivation}

\subsection{MoE Training at Scale}

The Mixture-of-Experts architecture has undergone a significant structural shift in recent years.
Table~\ref{tab:moe-models} surveys representative MoE models.
Early designs such as Mixtral~\cite{mixtral_of_experts} and DBRX~\cite{dbrx} adopt a coarse-grained architecture with a small number of large experts,
while more recent models such as DeepSeek-V3~\cite{deepseekv3} select 8 out of 256 fine-grained experts for each token.
This shift toward fine-grained MoE increases the number of activated experts and, 
critically, amplifies MoE-related activation memory,
making activation management a first-class concern for training MoE model at scale.

\begin{table}[h]
    \centering
    \vspace{-1em}
    \caption{Architecture of Recent MoE Models.}
    \vspace{-1em}
    \begin{tabular}{lcc}
    \toprule
    Models & Activated/Total Parameters & TopK/\#Experts \\
    \midrule
    Mixtral~\cite{mixtral_of_experts} & 12.9B/46.7B & 2/8 \\
    DBRX~\cite{dbrx} & 36B/132B & 1/4 \\
    GLM-4.5~\cite{glm4.5} & 32B/355B & 8/160 \\
    DeepSeek-v3~\cite{deepseekv3} & 37B/671B & 8/256 \\
    Kimi-k2~\cite{kimik2} & 32.6B/1040B & 8/384 \\
    \bottomrule
    \end{tabular}
    \vspace{-1em}
    \label{tab:moe-models}
\end{table}

Alongside this trend, the nature of expert load imbalance has also evolved.
Early MoE systems suffered from persistent, widespread imbalance across training iterations,
which motivated compute-focused mitigation strategies~\cite{fastmoe,bagualu}.
Advances in routing algorithms~\cite{gshard,wang2024auxiliary}
have largely eliminated this chronic imbalance.
However, imbalance has not disappeared: it now manifests as rare, high-magnitude spikes.
As shown in Figure~\ref{fig:main}(a), our measurements reveal that $97.1\%$ of training iterations
exhibit an imbalance rate below $1.2\times$, while $2.9\%$ of iterations exceed that rate with maximum $2.6\times$.
This shift in character changes the relevant system objective:
since spikes are infrequent, the cost of slowing down to handle them is low,
but an OOM failure, which requires manual intervention and job restart, is catastrophic.
Existing MoE training systems~\cite{smartmoe,fastermoe,bagualu,lina,flexmoe_nie} focus on compute load balancing.
They are poorly equipped to prevent the memory failures caused by occasional spikes.

To train fine-grained MoE model at scale,
mitigating activation memory becomes important.
The standard approach to mitigating activation memory draws on two families of techniques.
\textit{Recomputation} discards activations during the forward pass
and recomputes them from saved checkpoints during the backward pass, trading compute for memory~\cite{chen2016trainingdeepnetssublinear}.
\textit{Offloading} transfers activations to CPU memory during the forward pass
and prefetches them back to GPU before the corresponding backward computation~\cite{swapadvisor}.
Despite their differences, all existing systems share the same static workflow to apply activation mitigation:
(1) profile or model the memory footprint of a target configuration,
(2) select an activation mitigation strategy that fits within a fixed memory budget,
and (3) apply that strategy uniformly throughout training.
This workflow \textbf{implicitly assumes activation memory is stable across iterations}.
The occasional memory spikes of fine-grained MoE break this assumption:
a strategy sized for the average case will OOM on a spike,
while a strategy sized for the worst case wastes memory during the most of iterations.

\subsection{Dynamic Memory Pressure: A Quantitative Case}
\label{sec:motivation}

To quantify the cost of static activation management in MoE training,
we run a representative training workload on 512 GPUs using a 423B-A30B MoE model,
as shown in \Cref{fig:main}.
We use a state-of-the-art training system AdaPipe~\cite{adapipe},
and config its memory management policy as under-provisioning:
the activation memory budget is set for the average case,
maximizing memory utilization and MFU during normal iterations
but leaving no headroom to absorb spikes.
Specifically, selective recompute is applied for several operators.
This policy achieves a high average MFU of $30.0\%$ during normal iterations,
but encounters OOM failures in approximately $2.9\%$ of spike iterations,
each requiring a job restart that incurs a recovery overhead of $284$ seconds, which is $16\times$ the time cost of a normal iteration.
We profile the cold-start overhead and find most of the time is spent on large-scale distributed initialization and GPU kernel warmup,
which is hard to eliminate.
Note that this measurement excludes human response latency, which can further increase the total recovery time.
After restart casued by OOM,
the policy is switch to over-provisioning.
Specifically, full-layer recompute is applied so that activation memory consumption is not influenced by MoE imbalance rate.
This policy runs stably but delivers a sustained MFU of only $24.6\%$,
because the unused memory headroom forecloses more efficient configurations.

\rev{We argue from first principles that this spike is an inevitable consequence of the input token distribution, not an artifact of any particular routing algorithm, via a four-step chain.
\textit{(1) Routing randomness.} Fine-grained top-$k$ routing lets each token independently select its experts, introducing per-micro-batch sampling noise. Long-run load-balancing mechanisms --- an auxiliary loss~\cite{gshard} or a bias correction~\cite{wang2024auxiliary} --- cannot suppress this noise: each correction is computed from aggregate load observed over past iterations and fixed \emph{before} the current micro-batch is routed, so neither can condition on that batch's realized content.
\textit{(2) Token-distribution skew.} A micro-batch whose token distribution happens to be unfavorable --- a property of the data, not the algorithm --- therefore concentrates tokens on a small subset of experts regardless of the balancing mechanism in place, since any correction reacts to past load at least one iteration too late; our measurements show this occurs in a small fraction of micro-batches, with up to $2.6\times$ imbalance.
\textit{(3) Expert placement.} Fine-grained architectures pair a large expert count with a high expert-parallelism degree, so each device hosts only a few local experts; expert-level imbalance therefore translates directly into device-level \emph{token concentration}, since skew on locally-hosted experts accumulates activations on that device while skew on remote experts is carried away by all-to-all dispatch. This directionality is why a given skew's impact cannot be determined until the gate is computed (\S\ref{subsec:provisioning}).
\textit{(4) Warmup accumulation.} During the pipeline warmup phase (\Cref{fig:act-cost}c), multiple forward micro-batches stack without an intervening backward pass, so a single skewed batch can push the cumulative peak beyond device memory capacity.
The root cause is thus the finite-sample variance of per-token independent routing acting on a shifting data distribution --- an architectural property of any fine-grained MoE, not a correctable bug.}

\begin{figure}[t]
    \centering
    \begin{subfigure}{\linewidth}
        \centering
        \includegraphics[width=\linewidth]{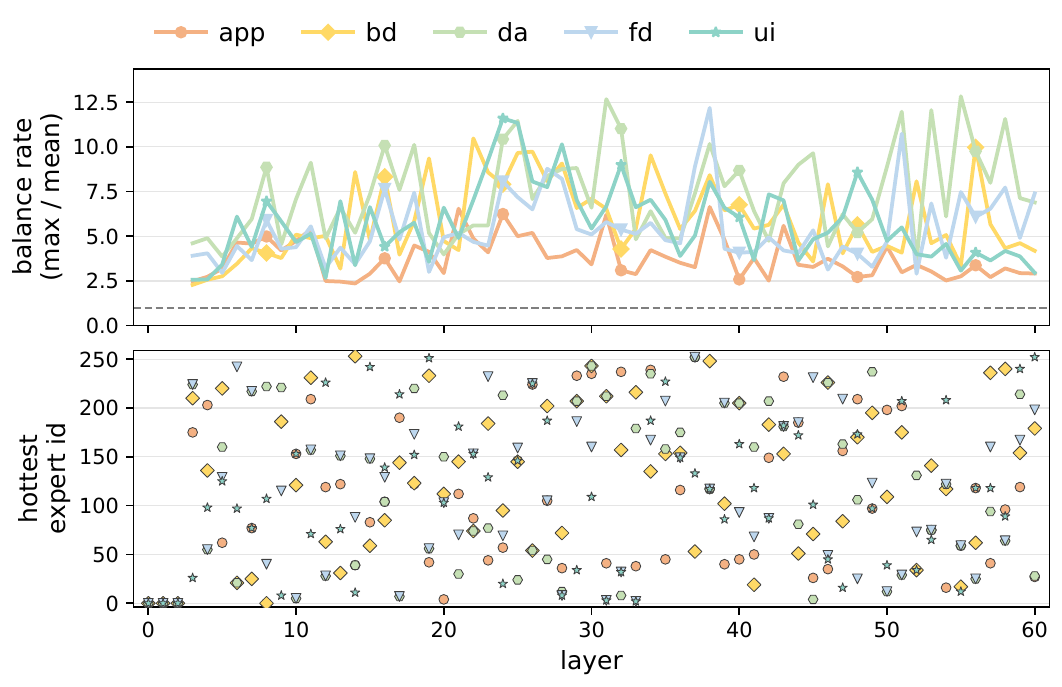}
        \caption{\rev{DeepSeek-V3.1-Terminus}}
    \end{subfigure}
    \begin{subfigure}{\linewidth}
        \centering
        \includegraphics[width=\linewidth]{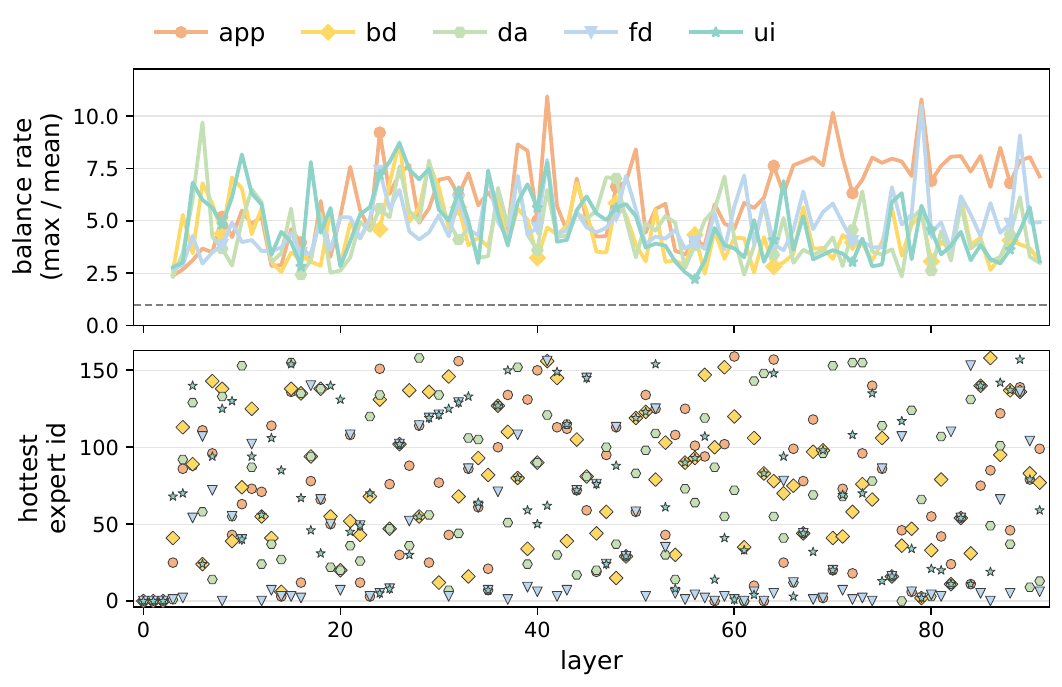}
        \caption{\rev{GLM-4.7}}
    \end{subfigure}
    \vspace{-1em}
    \caption{\rev{Expert-load skewness across SWE coding tasks.}}
    \vspace{-1em}
    \label{fig:domain-skew}
\end{figure}

\rev{We support this argument with direct evidence that real input data measurably alters expert-load skew, which by the same mechanism implies similar variation across training iterations.
We measure two deployed models with structurally different routing algorithms: DeepSeek-V3.1-Terminus (expert bias with a sequence-level auxiliary loss) and GLM-4.7 (expert bias with a sigmoid gate, no auxiliary loss).
For each model, we collect prefill-stage per-token routing decisions from software-engineering agent trajectories spanning five task domains --- application development (app), build and deployment (bd), data analysis (da), frontend development (fd), and UI optimization (ui) --- and compute, per layer and domain, the expert-load balance rate (max/mean, \Cref{fig:domain-skew} top) and the hottest expert (bottom).
\Cref{fig:domain-skew} shows that, for both models, the balance rate and even the hottest expert's identity differ substantially and systematically across domains at the same layer.
Since token-choice routing has no notion of task domain, this difference can only originate from the input token distribution: each domain induces a distinct mixture over the router's training content.
A training corpus is precisely such a heterogeneous, evolving mixture, so the same mechanism implies that the data mixture a given iteration samples determines that iteration's skew, explaining why spikes are sporadic rather than persistent.
This domain-dependent skew appears under both DeepSeek-V3.1-Terminus's auxiliary loss and GLM-4.7's sigmoid-gate correction, confirming empirically, beyond the argument above, that the root cause is intrinsic to token-choice routing rather than to any one balancing mechanism.}

These results expose the fundamental mismatch at the core of the problem:
existing systems apply a \emph{static} activation mitigation strategy to a training workload
whose activation pressure is inherently \emph{dynamic}.
This mismatch directly accounts for both the performance deficit of conservative approaches
and the stability failures of aggressive ones.
We need a system that dynamically adjusts its activation mitigation strategy
at runtime in response to actual memory pressure:
a capability that the static workflow of existing systems cannot provide.

\section{Overview}

\begin{figure}
    \centering
    \includegraphics[width=\linewidth]{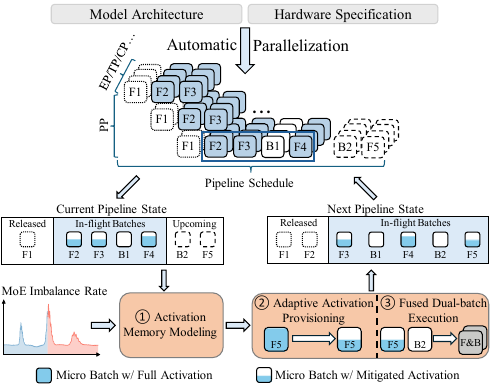}
    \caption{Overview of \sysname{}.}
    \vspace{-2em}
    \label{fig:bgl-overview}
\end{figure}

We propose \sysname{},
a training system that handles dynamic memory fluctuation in large-scale MoE model training.
\Cref{fig:bgl-overview} presents an overview.
Before training starts, \sysname{} follows existing practice
to generate an optimal hybrid-parallel strategy from training specification,
with the assumption of balanced MoE workload.
At runtime, \sysname{} performs \textit{flexible activation provisioning} to deal with imbalanced and unpredictable MoE workload.
At each pipeline scheduling point,
\sysname{} decides the execution plan (memory usage and operator execution order) for the next micro batch
with three steps: 

\noindent\textbf{Activation memory modeling} (\S\ref{sec:model}).
Given the current pipeline state and MoE imbalance rate,
\sysname{} evaluates the activation memory demand of the upcoming micro-batch
using the cost model developed in \S\ref{sec:model},
and assesses the cost-benefit tradeoff of each mitigation strategy.
\noindent\textbf{Adaptive Activation Provisioning} (\S\ref{subsec:provisioning}).
Then, \sysname{} applies diverse activation mitigation strategies for each batch.
A pipeline-aware memory budget is computed for the micro-batch,
and an iterative algorithm selects per-operator mitigation strategies
(recomputation, offloading, or quantization)
to keep the peak memory within GPU capacity
while minimizing performance overhead.
\noindent\textbf{Fused dual-batch execution} (\S\ref{subsec:overlap}).
Finally, \sysname{} examines the chosen mitigation strategies
of the current and adjacent micro-batches
and uses a lightweight solver to interleave operators' execution across computation and communication streams,
hiding the mitigation overhead in hardware bubbles.

These three steps collectively produce the execution plan
for the next pipeline state with optimized activation memory consumption,
minimizing total execution time without exceeding the device memory limit.


\section{Activation Memory Modeling}
\label{sec:model}

We now develop the \textit{activation cost model},
a unified performance model that jointly captures the time and memory cost
of activation management.
We first define the cost model from the operator level to a full iteration (\S\ref{subsec:cost-model}),
then quantify existing activation mitigation strategies under this model (\S\ref{sec:roofline}).

\subsection{Activation Cost Model}
\label{subsec:cost-model}

\begin{figure}[h]
    \centering
    \begin{subfigure}{\linewidth}
        \centering
        \includegraphics[width=\linewidth]{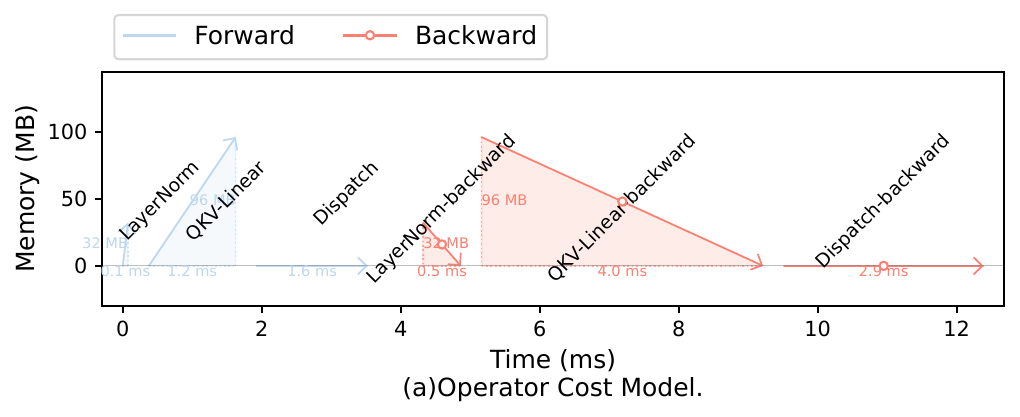}
    \end{subfigure}    
    \begin{subfigure}{\linewidth}
        \centering
        \includegraphics[width=\linewidth]{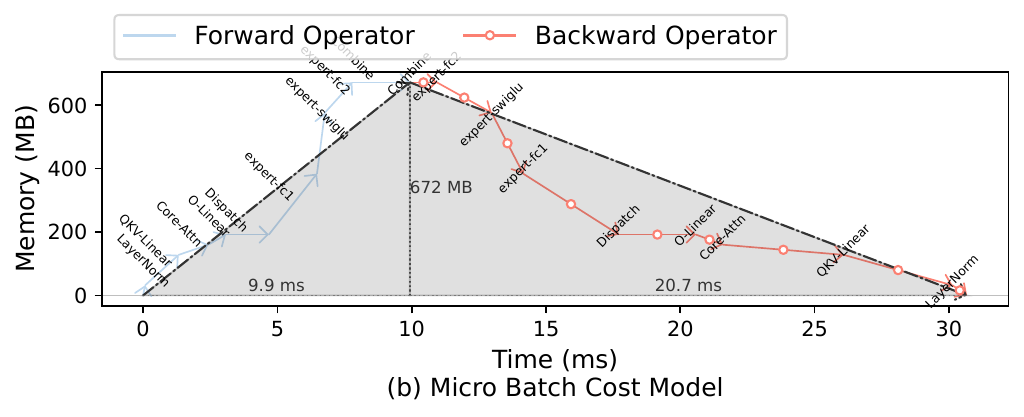}
    \end{subfigure}
    \begin{subfigure}{\linewidth}
        \centering
        \includegraphics[width=\linewidth]{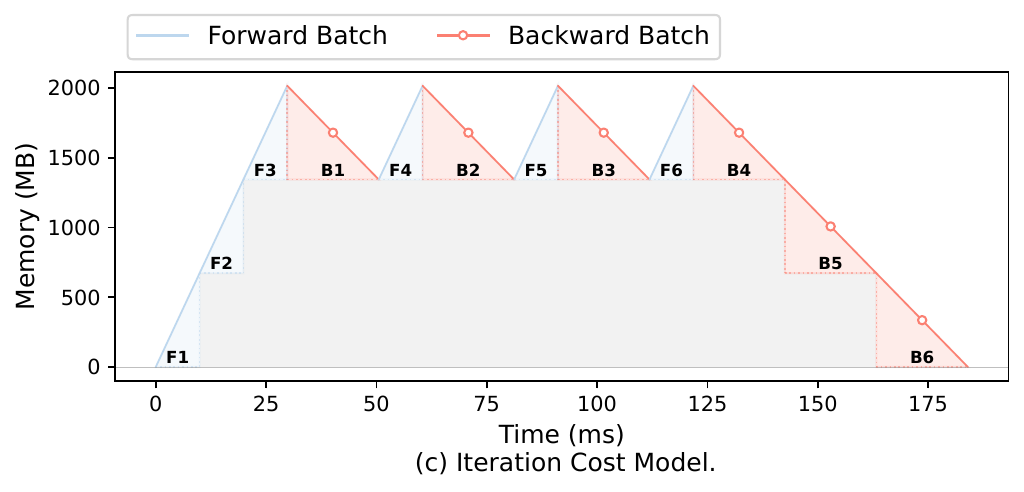}
    \end{subfigure}
    \vspace{-2em}
    \caption{Activation Cost Model at Different Levels.}
    \label{fig:act-cost}
    \vspace{-2em}
\end{figure}

\noindent\textbf{Operator activation cost (\Cref{fig:act-cost}a).}
We characterize each operator by an \textit{activation cost}:
a tuple $(\Delta t,\, \Delta m)$
where $\Delta t$ is the operator's execution time
and $\Delta m$ is the change in activation memory
(positive for forward allocation
and negative for backward release).
Visually, the activation cost traces a trajectory
on an \textit{activation memory curve}
whose $x$-axis is time and $y$-axis is GPU memory occupancy:
a forward operator corresponds to a right-triangle
with its hypotenuse rising to the upper-right,
and a backward operator corresponds to a right-triangle
with its hypotenuse falling to the lower-right
(Figure~\ref{fig:act-cost}).
As each operator corresponds to a single kernel launch,
the piecewise-linear approximation closely tracks the actual memory footprint.
Operators in an MoE transformer block show very different characteristics.
We illustrate three representative operators: LayerNorm, Linear, and MoE dispatch communication.
Comparing LayerNorm and Linear,
the ratio of $\Delta m$ and $\Delta t$ has a significant difference. 
It leads to different mitigation bandwidth introduced in the next section.
And for dispatch communication,
there is no memory cost because it has no trainable parameters.

\noindent\textbf{Micro-batch activation cost (\Cref{fig:act-cost}b).}
A forward or backward micro-batch
is a sequence of operators executed on a single pipeline stage.
Its activation cost is the concatenation
of its constituent operators' costs,
yielding an approximated composite triangle on the activation memory curve
with total height $\sum \Delta m$ and total width $\sum \Delta t$.
Since every operator only contributes a small portion of the total activation memory, 
this approximation is acceptable.
For simplicity,
we will use this approximation in the following analysis on micro-batch and iteration level modeling.

\noindent\textbf{Iteration activation cost (\Cref{fig:act-cost}c).}
Under pipeline parallelism,
a training iteration executes multiple micro-batches
across three pipeline scheduling phases.
On the activation memory curve,
these phases manifest as characteristic patterns
of micro-batch cost triangles:
\textit{Warmup.}
  only forward batches execute.
  Forward triangles stack sequentially,
  and the memory curve rises monotonically
  as each micro-batch's activations accumulate.
\textit{Steady state (1F1B).}
  forward and backward batches interleaved.
  Each forward triangle (rising) is immediately followed
  by a backward triangle (falling),
  so the memory curve oscillates around a plateau.
\textit{Cooldown.}
  only backward batches execute.
  Backward triangles stack sequentially,
  and the memory curve descends
  as activations are progressively deallocated.

Iteration-level cost model uncovers the formulation of an activation memory peak.
During MoE model training,
each micro batch incurs different MoE imbalance workloads,
resulting in varied activation memory consumption.
The warmup phase's forward batches stack the most part of consumption,
then the steady phase's forward batches will exceed the device memory limit if encountering severe MoE imbalancing.
Once into cooldown phase,
it is impossible to encounter the out-of-memory problem, as the memory peak has passed.

\begin{figure}[h]
    \centering
    \begin{subfigure}{\linewidth}
        \centering
        \includegraphics[width=\linewidth]{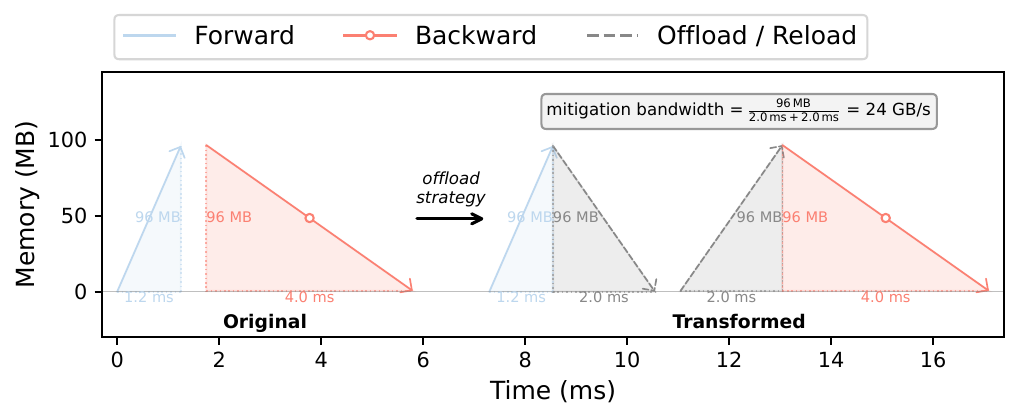}
    \end{subfigure}
    \begin{subfigure}{\linewidth}
        \centering
        \includegraphics[width=\linewidth]{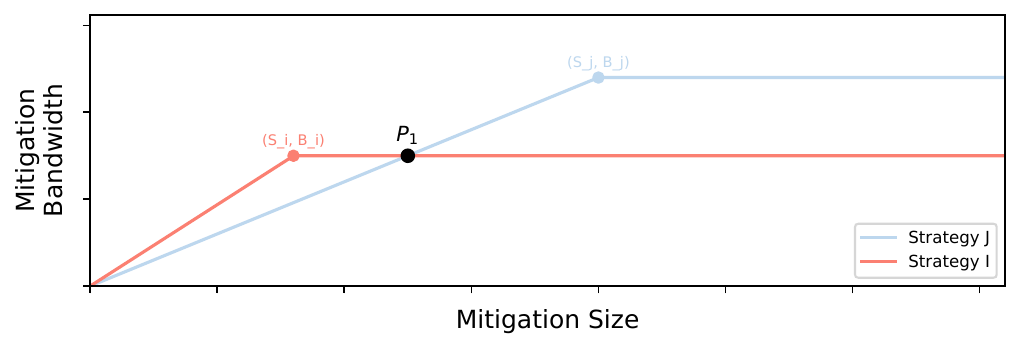}
    \end{subfigure}
    \vspace{-2em}
    \caption{Mitigation Strategy Modeling.}
    \vspace{-2em}
    \label{fig:act-cost-transform}
\end{figure}

\begin{figure*}[t]
    \centering
    \includegraphics[width=\linewidth]{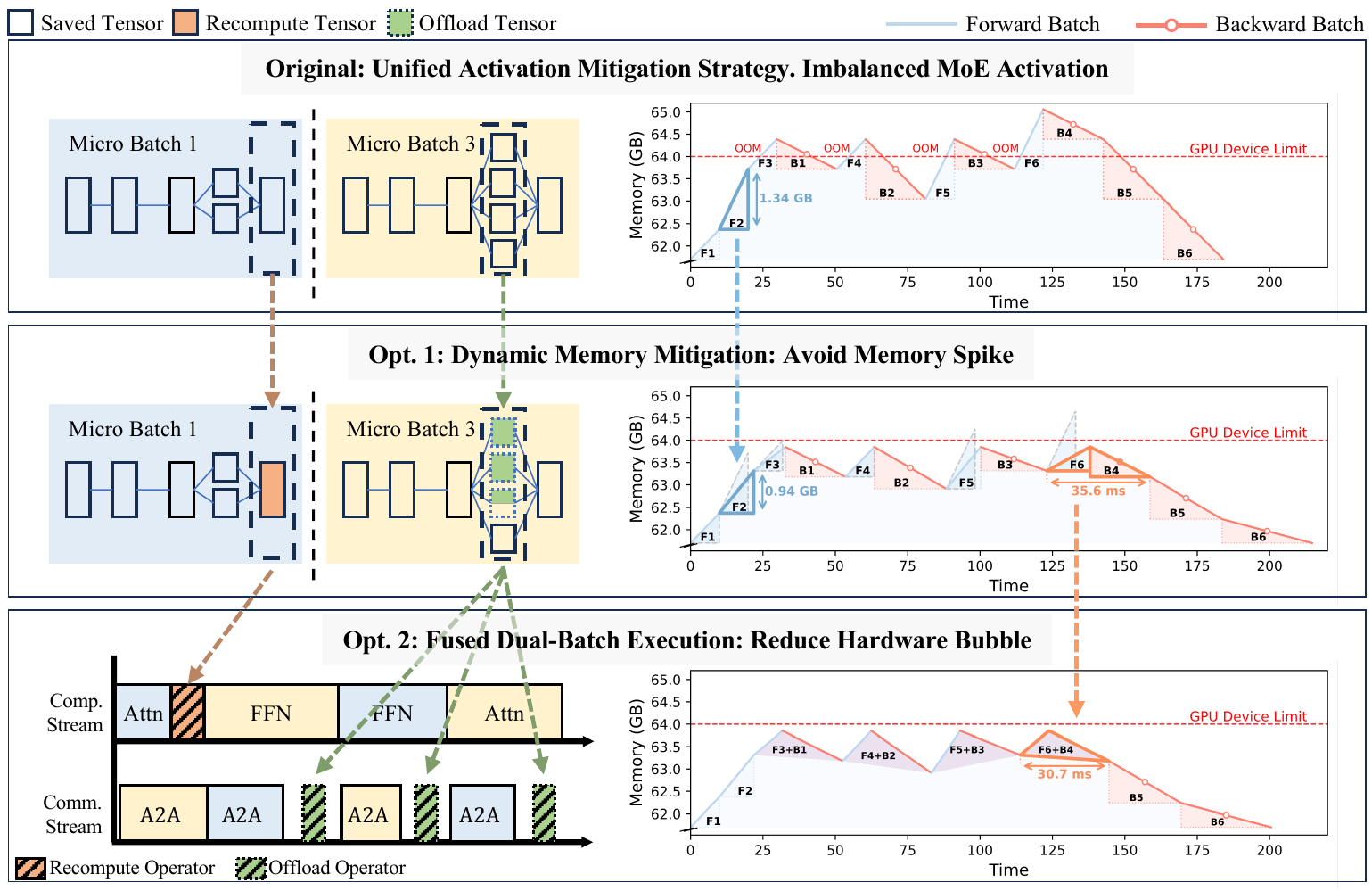}
    \caption{\sysname{} Runtime Workflow.}
    \vspace{-2em}
    \label{fig:overview}
\end{figure*}

\subsection{Activation Mitigation Cost}
\label{sec:roofline}

\noindent\textbf{Mitigation Strategies}.
Existing activation mitigation strategies
correspond to transformations of the activation cost tuple.
We summarize three strategies:
%
\textbf{Recomputation} discards the forward activation
  and regenerates it during the backward pass.
  This reduces $\Delta m$ of the forward operator (less memory allocated)
  at the expense of increasing $\Delta t$ of the backward operator
  (which must recompute before consuming the activation).
\textbf{Offloading} transfers the activation to CPU memory
  during the forward pass and reloads it before the backward pass.
\textbf{Quantization} casts the activation to a lower precision (e.g., BF16 to FP8)
  for storage and casts it back before use.
  Both reduce $\Delta m$ of the forward operator
  while increasing $\Delta t$ of both the forward operator
  (offload or cast overhead) and the backward operator (reload or cast-back).

\Cref{fig:act-cost-transform} shows an example cost model of offloading strategy.
Each transformation trades memory savings
for additional time cost ---
a tradeoff that the cost model makes explicit and quantifiable.
Each mitigation strategy transforms the activation cost tuple
of an operator in a characteristic way.
To compare these strategies, we adopt a \textbf{roofline-style model}.
For each strategy, we define its \textit{mitigation bandwidth}
as the rate at which it can reduce activation memory
($\text{GB saved} / \text{sec. overhead}$),
plotted against the target activation size.
As shown in \Cref{fig:act-cost-transform},
the resulting curve exhibits a roofline shape:
below a threshold size $S_i$, the bandwidth is limited by launch overhead;
above $S_i$, it saturates at a peak bandwidth $B_i$.
The knee point $(S_i, B_i)$ is determined by
the hardware characteristics and the operator type.
Given two strategies with knee points $(S_i, B_i)$ and $(S_j, B_j)$
where $B_i < B_j$,
strategy $j$ is more efficient for large activations.
For small activations, strategy $i$ can be preferable
when $S_i < S_j$ and the activation size
falls below the crossover point $P_1$, where mitigation size is $S_i \cdot B_j / B_i$.
%
Given a certain operator and input data size,
the roofline model guides to find the time-minimizing mitigation strategy.
In practice,
compute-bound operators
usually have a roofline favoring offloading strategy;
while memory-bound operators, such as layer normalization,
have a roofline favoring recomputation strategy.



\section{Model-Guided Runtime Memory Provisioning}

\noindent\textbf{Problem Formulation.}
For dense models,
every micro-batch produces the same activation cost,
so the activation memory curve is identical across batches and iterations.
For example, in \Cref{fig:act-cost}c, 
the peak activation memory starts at the end of the warmup phase
and sustains until the beginning of the cooldown phase. 
A static mitigation strategy planned before training
is sufficient to prevent out-of-memory errors.

MoE models break this assumption.
The dynamic token-to-expert routing causes
the $\Delta m$ of MoE-related forward operators
to fluctuate with the imbalance rate:
when more tokens are routed to locally-hosted experts,
the activation footprint grows.
As shown in \S\ref{sec:motivation},
this fluctuation is unpredictable and can spike by $1.5\times$ or more,
making any fixed strategy either wasteful in the common case
or unsafe during spikes.
Given a cluster and an MoE model, 
the system must decide, for each operator in each micro-batch,
which activation mitigation strategy to apply.
The optimization objective is:
\[
  \min \sum \Delta t \quad \text{s.t.} \quad
  \text{peak memory} \le M_{\max},
\]
where $M_{\max}$ is the GPU memory capacity
and the sum is over all operators in the iteration.
In other words, the system minimizes the total execution time
(i.e., the total width of the activation memory curve)
subject to keeping the curve's peak height within GPU capacity.
Because the activation costs of MoE operators
are only known at training runtime
(once the imbalance rate is observed),
this optimization must be solved \emph{online},
during each iteration's pipeline scheduling points.

\noindent\textbf{Runtime Workflow.}
At a pipeline scheduling point,
\sysname{} obtains the activation memory cost of current batches with the activation cost model (\S\ref{sec:model}).
Then, \sysname{} employs two complementary optimizations
that address the two dimensions of the activation cost tuple respectively.
\textit{Dynamic Activation Memory Mitigation} (\S\ref{subsec:provisioning})
selects per-operator activation mitigation strategy
to keep the peak memory within the GPU capacity,
taming the \textit{height} of the activation memory footprint curve.
\textit{Fused Dual-Batch Execution} (\S\ref{subsec:overlap})
interleaves adjacent forward and backward micro-batch operations
to reduce the total execution time,
compressing the \textit{width} of the curve.

\Cref{fig:overview} illustrates the effect of the two optimizations
using the activation memory curve of a training iteration.
The naive curve exceeds GPU capacity at several micro-batches.
After applying dynamic activation memory mitigation, the peak is brought within capacity,
but the added recomputation and offload operations increase the total execution time.
After dual-batch fused execution, the timespan recovers
by hiding mitigation operations behind useful computation.

\subsection{Dynamic Activation Memory Mitigation}
\label{subsec:provisioning}

The goal of runtime activation provisioning is to select,
for each operator in each micro-batch,
a mitigation strategy (none, recompute, offload, or quantize)
such that the peak activation memory stays within GPU capacity
while minimizing the total time overhead.
In this section, we address four questions:
how to deal with unpredictable MoE spikes,
how much memory each micro-batch can afford considered the unpredictable workload,
which mitigation strategy to apply given the budget,
and how to handle conflicts when composing strategies.


\noindent\textbf{Deferred Planning for Unpredictable MoE Spikes.}
The actual MoE imbalance workload is unpredictable until the execution of MoE gate operator within the forward micro batch.
Therefore, it is impossible to exactly preserve the activation memory quota for an incoming forward batch.
However, we find that deferring activation memory provisioning to the next micro batch is acceptable, because in \S\ref{sec:model} the iteration-level activation cost model reveals that the activation memory peak is compounded by multiple batches.
During the pipeline warmup phase,
if current batch unexpectedly consumes too much activation memory,
there is still time to avoid out-of-memory by
reducing the memory quota of the next batch.
During the pipeline steady phase,
if current batch meets an unexpected high MoE activation memory peak,
\sysname{} could downgrade to full recompute of MoE module once the MoE gate is computed yet experts are not, 
thus avoiding that peak.
This conception of deferred planning is further quantified in the following memory budget assignment algorithm.

\noindent\textbf{Pipeline-Aware Memory Budget Assignment.}
The activation memory budget depends on both the pipeline schedule phase
and the actual memory pressure at runtime.
Algorithm~\ref{alg:budget} computes a per-micro-batch budget
that guarantees all concurrent micro-batches fit within GPU memory.
Let $B$ denote the total number of micro-batches,
$B_w$ the number of warmup batches,
$M_{\max}$ the GPU memory capacity,
and $M_0$ the minimum per-batch activation memory
(achieved under full recomputation, retaining only the block input tensor).
During warmup (lines 3--9), no backward pass releases memory,
so the budget for each new micro-batch is computed by
evenly distributing the remaining capacity among all yet-to-be-launched batches,
while reserving $M_0$ for each to guarantee feasibility.
During the steady phase (lines 10--16), each forward is preceded by a backward
that frees the activations of an earlier micro-batch (line 11),
and the freed memory is fully allocated to the incoming batch (line 12).
In both phases, the budget is updated to the actual memory consumed (lines 7, 14),
so that over-estimation in one batch does not cascade into under-allocation for subsequent batches.

\noindent\textbf{Roofline-Guided Iterative Activation Mitigation.}
Given the memory budget for a micro-batch,
the system must select which operators to mitigate
and with which strategy.
In terms of the activation cost model,
this is choosing a mitigation strategy for each operator
such that the total $\sum \Delta m$ (the micro-batch triangle height)
stays within the budget
while minimizing the total $\sum \Delta t$ (the micro-batch triangle width).
We solve this with a greedy iterative algorithm (Algorithm~\ref{alg:transform}).
Using the roofline model
of individual mitigation strategy characterized in (\S\ref{sec:roofline}),
every operator has a candidate mitigation strategy.
At each iteration, we evaluate the memory reduction
of every candidate single-operator mitigation on the current configuration,
select the one with the largest reduction,
verify that it does not conflict with previously applied strategies
(detailed below),
and apply it if valid.
The algorithm terminates when the memory budget is satisfied.
The potential memory reduction is re-evaluated at each iteration,
since applying one strategy may change the benefit of subsequent ones
(e.g., recomputing a set of operators makes offloading its intermediate nodes unnecessary).

\begin{algorithm}
\caption{Pipeline-Aware Memory Budget Assignment}
\label{alg:budget}
\begin{algorithmic}[1]
\Require Total micro batches $B$, warmup count $B_w$,
         GPU capacity $M_{\max}$, min budget $M_0$

\State $\textit{budget}[0 \ldots B{-}1] \gets 0 $
\State $M_{\text{used}} \gets$ non-activation memory usage
\For{$i = 0$ \textbf{to} $B_w - 1$} \Comment{Warmup}
  \State $r \gets B_w - i$
  \State $\textit{budget}[i] \gets (M_{\max} - M_{\text{used}} - r \cdot M_0) / r + M_0$
  \State $\textit{actual} \gets \Call{RunForward}{i, \textit{budget}[i]}$
  \State $\textit{budget}[i] \gets \textit{actual}$
  \State $M_{\text{used}} \mathrel{+}= \textit{actual}$
\EndFor
\For{$i = B_w$ \textbf{to} $B - 1$} \Comment{Steady}
  \State $M_{\text{used}} \mathrel{-}= \textit{budget}[i - B_w]$
  \State $\textit{budget}[i] \gets M_{\max} - M_{\text{used}}$
  \State $\textit{actual} \gets \Call{RunForward}{i, \textit{budget}[i]}$
  \State $\textit{budget}[i] \gets \textit{actual}$
  \State $M_{\text{used}} \mathrel{+}= \textit{actual}$
\EndFor
\end{algorithmic}
\end{algorithm}

\begin{algorithm}
\caption{Iterative Activation Mitigation}
\label{alg:transform}
\begin{algorithmic}[1]
\Require Set of operator activation costs $\{(\Delta t_i, \Delta m_i)\}$, memory budget $M$
\Ensure Mitigation with $\sum \Delta m \leq M$
\State $\textit{mem} \gets \sum \Delta m_i$
\If{$\textit{mem} \leq M$}
  \State \Return \Comment{No strategy needed}
\EndIf
\While{$\textit{mem} > M$}
  \State $t^* \gets$ available strategy with largest memory reduction
  \If{\Call{Conflict}{$t^*$}}
    \State Mark $t^*$ as unavailable; \textbf{continue}
  \EndIf
  \State Apply $t^*$: update $(\Delta t_i, \Delta m_i)$ of affected operators
  \State $\textit{mem} \gets \sum \Delta m_i$
\EndWhile
\end{algorithmic}
\end{algorithm}

\noindent\textbf{Heterogeneous Mitigation Conflicts.}
Existing systems typically employ mitigation strategies in isolation
and do not encounter conflicts between them.
When composing all three types of strategies,
we find that independently valid activation mitigation can interact to produce redundant storage,
a phenomenon we term \textbf{conflicting mitigation}.

The root cause is that a single logical activation tensor may serve as input
to multiple downstream operators.
When different strategies are applied to different consumers of the same tensor,
inconsistencies arise.
We identify two common patterns:
\textit{Case 1: Multi-precision conflict.}
The same tensor is transformed into different precisions at different consumers,
causing multiple copies to coexist.
For example, the MoE layer input serves as an input to both the MoE router
(which typically requires FP32 for numerical stability)
and the shared expert (which typically uses BF16 or FP8).
If quantization is applied independently to each consumer,
both an FP32 and an FP8 copy are stored,
increasing rather than reducing total memory.
\textit{Case 2: Multi-device conflict.}
The same tensor is offloaded at one consumer but remains on the GPU at another,
so the GPU copy cannot be freed.
For example, 
the core attention output tensor serves as the activation tensor for core attention operator and the input tensor for the following out-projection.
If offloading is applied to the attention output
but the out-projection saves the same tensor on GPU,
the GPU tensor persists, and the offload yields no memory savings.

To detect these conflicts, we perform a check after each candidate strategy.
The check traces each optimized operator back to its source activation tensor
and verifies that all consumers of the same tensor
have consistent precision and device placement.
If a conflict is detected, the candidate strategy is rejected.
Since the algorithm selects the strategy in decreasing order of memory benefit,
a conflicting candidate always conflicts with a previously applied strategy
that yields a larger reduction;
skipping the current candidate is therefore always preferable
to reverting the earlier, more beneficial one.

\subsection{Fused Dual-Batch Execution}
\label{subsec:overlap}

\noindent\rev{\textbf{Motivation.}}
After dynamic activation memory mitigation ensures that the activation memory curve
stays within GPU capacity,
this step optimizes total execution time.
The applied mitigation strategies increase the total execution time,
because extra operators are introduced to delay the original operator execution.
To address this problem,
we analyze each pipeline scheduling point and find optimization opportunities when reordering the operators of adjacent micro batches is possible.
During the steady phase of pipeline parallelism,
each GPU simultaneously holds four types of operators:
a set of forward operators $F$ for the current forward batch,
a set of backward operators $B$ for the current backward batch,
a set of activation offloading operators $O$ for the previous forward batch,
and a set of activation recovery (i.e., reload and recomputation) operators $R$ for the next backward batch.
The data dependencies among the same type of operators form a simple chain,
while there are no data dependencies across different types, because they belong to different batches.
There is an opportunity to refine execution orders from different types 
to achieve better overlapping between computation and communication streams.
The optimization goal is to minimize overall execution time by planning an execution order of four types of operators.

\noindent\rev{\textbf{Formulation.}}
We formulate this planning problem as a dynamic programming (DP) problem.
Let state $F[i][j][k][l]$ be the minimum execution time of the first $i$ forward operators, the first $j$ backward operators, the first $k$ offloading operators and the first $l$ recovery operators.
The DP transfer equation can be formulated as follows:
\[
  F[i][j][k][l] = \min_{\substack{w \le i,\, x \le j\\ y \le k,\, z \le l}}
  \bigl( F[w][x][y][z] + T(w{:}i, x{:}j, y{:}k, z{:}l) \bigr)
\]

\noindent where T represents the direct overlapped execution time of consecutive operators from each type.
When updating F,
the optimal transfer points for DP states are recorded,
so that the optimal execution order can be constructed.
The number of DP states is $O(|F||B||O||R|)$, and each state has $O(|F||B||O||R|)$ transfer points.
So the overall time complexity of this DP problem is $O(|F|^2|B|^2|O|^2|R|^2)$.

\noindent\rev{\textbf{Runtime Solving.}}
Naively solving this DP problem could be expensive,
as a single micro batch typically consumes only tens of milliseconds.
Thus, we propose two optimizations to reduce the number of DP states and eliminate transfer costs:
%
First, we observe that DP states could be compacted.
There are many consecutive computation operators in a transformer block,
splitting them into different states cannot provide more overlapping opportunities to reduce execution time.
Therefore, we divide the original set $F$ into several stages, splitting by the communication operators.
And operators in the same stage are compacted into a single DP state.
For example,
an MoE transformer forward batch has more than $20$ operators but only $4$ stages,
as there are only two all-to-all communication operators used by expert parallelism.
This compaction strategy significantly reduces the number of DP states.
Second, we find opportunities for memorizing the optimal DP transfer points.
The operator sets are fully determined by the micro-batch memory budget described in the previous section,
therefore we could store DP states keyed by a discretized budget value.
Moreover,
we find the memory budgets exhibit strong temporal locality.
The pipeline phase repeats every $B_w$ micro-batches,
and MoE imbalance rates are stable for the majority of iterations.
Most iterations hit the cache and skip algorithms entirely,
reducing the amortized overhead to nearly zero.

\noindent\rev{\textbf{Correctness.}}
\rev{
Any such reordering is always valid, since an incorrect interleaving would silently corrupt training.
}
\textbf{Intra-class order is preserved by construction.}
The DP state $F[i][j][k][l]$ is defined over \emph{prefixes} of each class: index $i$ always denotes exactly the first $i$ forward operators in their original program order, and likewise for $B$, $O$, $R$.
Every transition in the recurrence above extends these prefixes ($i \to i'$ with $i' \ge i$, never $i' < i$); no transition permutes operators within a class.
Since the operators within one micro-batch's forward (or backward) pass form a linear producer-consumer chain, preserving prefix order exactly preserves every intra-class data dependency.
\textbf{Inter-class independence follows from the pipeline schedule.}
Denote a scheduling point by $F_xB_y$, where batch $x$ is the currently-dispatched forward and batch $y$ is the currently-dispatched backward (always $x \neq y$, since a batch's own forward and backward are never concurrent); $O$ then corresponds to batch $(x{-}1)$'s offload (the previous forward batch), and $R$ to batch $(y{+}1)$'s recovery (the next backward batch).
A sync barrier separates consecutive scheduling points, guaranteeing that every operator of $F_{x-1}B_{y-1}$ completes before any operator of $F_xB_y$ begins.
This barrier resolves the dependencies of $O$ and $R$ outside the current window, prior to any interleaving decision by the DP: $O$ consumes batch $(x{-}1)$'s activation, produced by $F_{x-1}$ at the \emph{preceding} scheduling point and therefore already complete by the barrier; $R$ produces the activation consumed by $B_{y+1}$ at the \emph{following} scheduling point $F_{x+1}B_{y+1}$, and is guaranteed complete by the next barrier before $B_{y+1}$ begins.
For instance, at a scheduling point with $x - y = 3$, e.g.\ $F_5B_2$: $F$ corresponds to batch 5, $B$ to batch 2, $O$ to batch 4, and $R$ to batch 3 --- four distinct batches, so the barrier argument above applies directly.
\sysname{} requires $x - y > 2$, equivalently a pipeline-parallel degree exceeding two: at $x - y = 2$, $O$'s target batch $(x{-}1)$ would coincide with $R$'s target batch $(y{+}1)$.
Every configuration in our evaluation uses a pipeline-parallel degree of at least $4$ (\Cref{tab:models}), and any training run spanning thousands of GPUs requires substantial pipeline depth to remain communication-efficient, so this boundary case does not arise in the regime this paper targets.

\begin{table*}[t]
\centering
\small
\caption{MoE model configurations.}
\vspace{-.5em}
\label{tab:models}
\begin{tabular}{ccccccccccccc}
\toprule
\textbf{Model} & \textbf{Hidden} & \textbf{MoE Hidden} & \textbf{\#Experts} & \textbf{Top-$K$} & \textbf{Attn. Head Dim} & \textbf{\#GPUs} & \textbf{DP} & \textbf{TP} & \textbf{EP} & \textbf{PP} & \textbf{VP} \\
\midrule
GLM-188B-A20B & 7168 & 4096 & 64 & 4 & 112 & 128 & 16 & 2 & 32 & 4 & 2  \\
GLM-319B-A30B & 8192 & 4096 & 96 & 6 &  128 & 256 & 16 & 2 & 32 & 8 & 1 \\
GLM-423B-A30B  & 8192 & 4096 & 128 & 6 &  128 & 512 & 32 & 2 & 64 & 8 & 2  \\
GLM-629B-A30B  & 8192 & 4096 & 192 & 6 &  128 & 1024 & 64 & 2 & 64 & 8 & 2  \\
GLM-783B-A35B  & 8192 & 5120 & 192 & 6 & 128 & 2048 & 128 & 2 & 64 & 8 & 1 \\
GLM-835B-A30B  & 8192 & 4096 & 256 & 6 &  128 & 4096 & 256 & 2 & 128 & 8 & 2  \\
GLM-1041B-A35B  & 8192 & 5120 & 256 & 6 & 128 & 8192 & 512 & 2 & 128 & 8 & 2  \\
GLM-1041B-A35B  & 8192 & 5120 & 256 & 6 & 128 & 12288 & 768 & 2 & 128 & 8 & 2  \\
\bottomrule
\end{tabular}
\vspace{-1.5em}
\end{table*}

\section{Evaluation}
\label{sec:eval}

\subsection{Setup}
\label{subsec:eval-setup}

\noindent\textbf{Hardware.}
All evaluation is conducted on a cluster with our vendor's GPU.
The GPU has $480$ TFLOPS peak performance of BF16 computation and $64$GB HBM.
Each compute node has 8 GPUs, where intra-node GPU bandwidth is $400$ GB/s and inter-node bandwidth is $400Gbps \times4$.
There are $1536$ nodes, in total $12288$ GPUs.

\noindent\textbf{Models.}
We evaluate on a family of MoE models at varying scales.
Table~\ref{tab:models} lists the hyperparameters (all models consist of $32$ layers).
We derive our model configurations from the GLM model family,
adjusting architecture parameters to align with the hardware characteristics
and maximize hardware utilization ---
a standard practice of model--cluster co-design
adopted across the industry~\cite{deepseekv3,kimik2}.

\noindent\textbf{Parallelism.}
At each scale, we use an automatic parallelization algorithm~\cite{smartmoe} to search hybrid parallelism strategies
under the assumption of balanced MoE workload.
For pipeline parallelism, interleaved 1F1B scheduling is used:
the ``VP'' column in Table~\ref{tab:models} denotes the number of layers per virtual pipeline stage.
For a fair comparison, all systems share the same offline-optimal parallelism configuration at the start of training.
The difference among systems therefore lies in
how they manage activation memory under this shared configuration,
which leads to different memory pressures and ultimately
different training stability and performance.

\noindent\textbf{Baselines.}
We compare against two state-of-the-art training systems.
\emph{Megatron-Kuai}~\cite{megatron-kuai} is a training system
for long-context dense models
that features fine-grained activation recomputation and offloading.
Its memory model does not account for MoE expert activations,
so we extend its memory modeling with MoE-specific terms
and evaluate its recomputation and offloading modules
using its open-source implementation.
\emph{AdaPipe}~\cite{adapipe} is a training system for memory-efficient pipeline parallelism.
It features adaptive recomputation and pipeline stage partitioning.
Its optimization only targets dense models,
and we find that adaptive pipeline partitioning is ineffective in the MoE training scenario.
Only the adaptive recomputation part of AdaPipe is enabled in our evaluation.

As discussed in \S\ref{sec:introduction},
both baseline systems manage memory statically.
We initialize each baseline with a memory strategy
configured for balanced MoE load. 
When an out-of-memory crash occurs,
we adjust the reduction strategy according to the observed imbalance rate
and manually restart training.

We report overall training throughput,
which averages per-iteration MFU over the entire run
including all restart overheads.
When computing the restart overhead,
we only count the machine warm-up latency
and exclude the human-in-the-loop time required for diagnosis and intervention.
The reported restart overhead is therefore a lower bound;
in production, the penalty would be strictly larger.

\subsection{End-to-End Performance}
\label{subsec:eval-e2e}

\begin{figure}[h]
    \centering
    \vspace{-1.5em}
    \includegraphics[width=\linewidth]{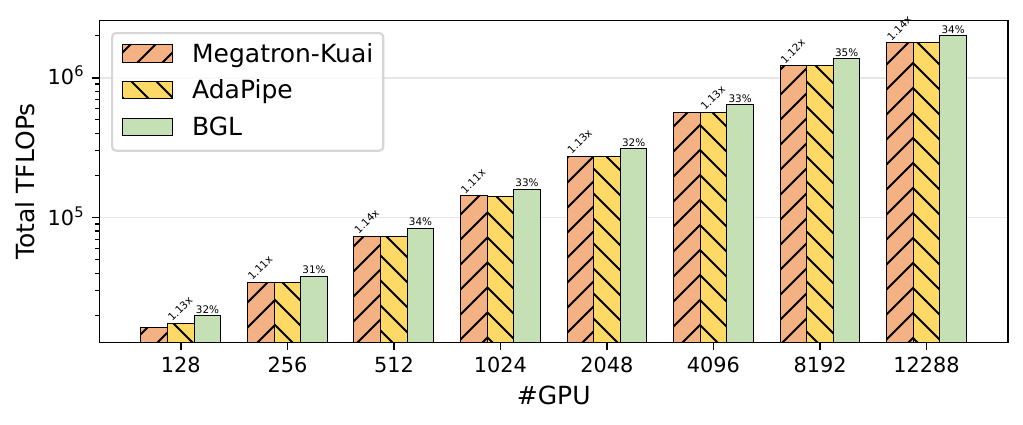}
    \vspace{-2em}
    \caption{End-to-end training throughput up to 12{,}288 GPUs.
Bar height indicates total TFLOPS (log scale);
labels above \sysname{} bars show its MFU,
labels above baseline bars show the speedup of \sysname{} over the better baseline.}
    \label{fig:end-to-end}
\end{figure}

We compare end-to-end training performance across cluster scales
ranging from 128 to 12{,}288 GPUs.
Figure~\ref{fig:end-to-end} shows the total training throughput (TFLOPS)
and per-GPU MFU for each system.
\sysname{} consistently outperforms both baselines at every scale.
Across all configurations, we achieve
an average speedup of $1.14\times$ over Megatron-Kuai
and $1.13\times$ over AdaPipe.
The average MFU of \sysname{} across all scales is $32.9\%$.
At 12{,}288 GPUs, \sysname{} sustains $2.012$ EFLOPs in total,
demonstrating that our system could scale
to over ten thousand GPUs without performance degradation.

\begin{figure}
    \centering
    \includegraphics[width=\linewidth]{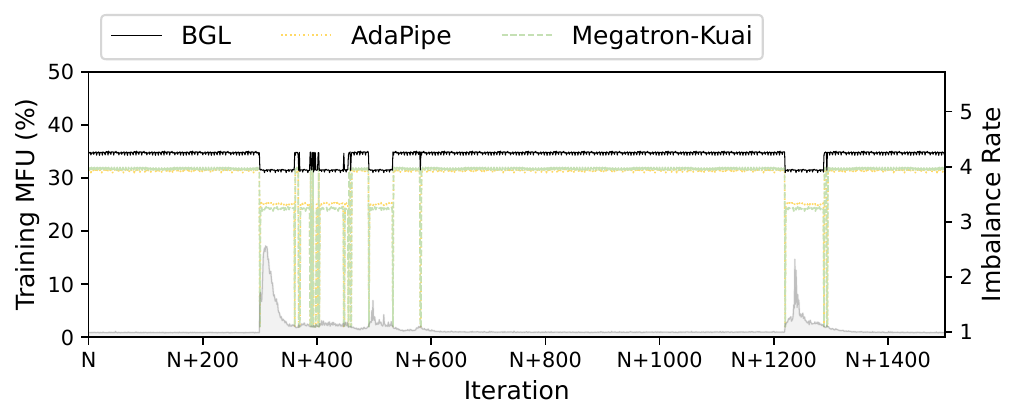}
    \vspace{-1.5em}
    \caption{Per-iteration MFU (left axis) and token-to-expert imbalance rate (right axis)
    over GLM-423B-A30B training.
    \sysname{} maintains stable MFU through imbalance spikes
    that cause significant performance degradation in baselines.}
    \label{fig:imbalance-mfu}
\end{figure}

\begin{figure}[t]
    \centering
    \begin{subfigure}{\linewidth}
        \centering
        \includegraphics[width=\linewidth]{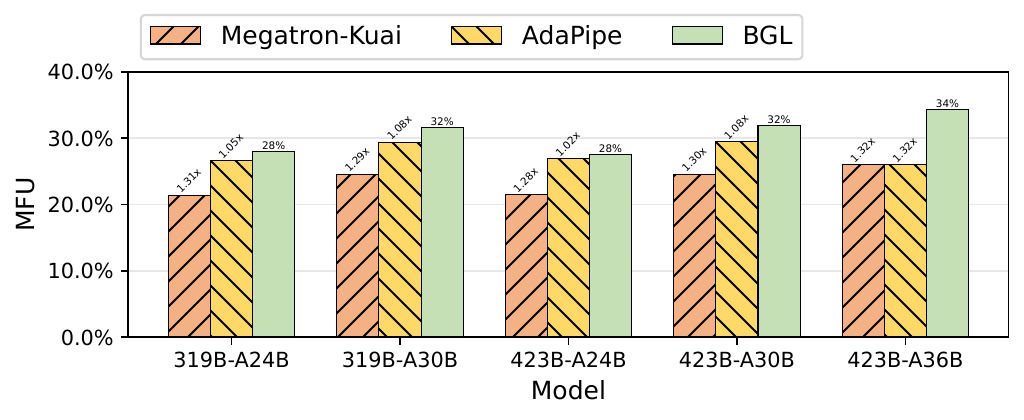}
    \end{subfigure}
    \begin{subfigure}{\linewidth}
        \centering
        \includegraphics[width=\linewidth]{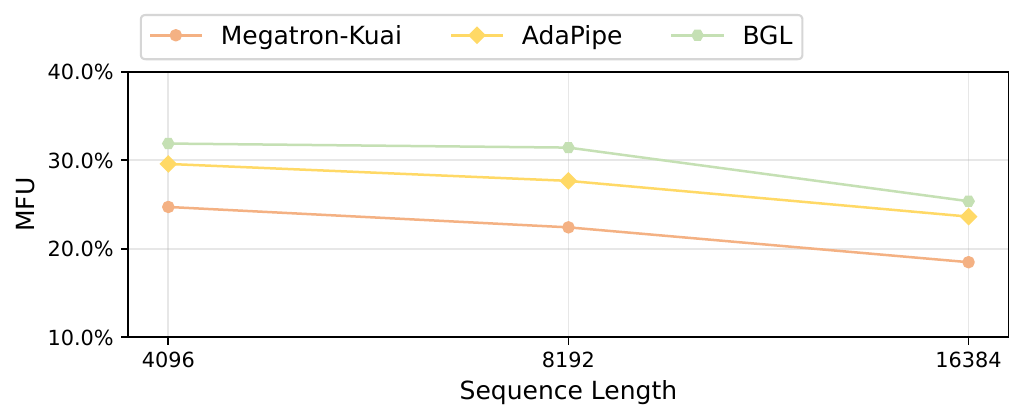}
    \end{subfigure}
    \vspace{-1.5em}
    \caption{MFU under varing configurations. 
    }
    \label{fig:config-study}
    \vspace{-1em}
\end{figure}



\subsection{Robustness under Varying Memory Pressure}
\label{subsec:eval-robustness}


We now evaluate whether \sysname{}'s advantage holds
when the nature of memory fluctuation changes ---
both within a single training run (dynamic variation)
and across workload configurations (static variation).

\noindent\textbf{Runtime Adaptivity to Imbalance Spikes.}
We select a representative 423B (A30B) model training on 512 GPUs
and trace per-iteration MFU alongside the MoE imbalance rate.
Figure~\ref{fig:imbalance-mfu} shows the result.
We find that imbalanced MoE token-to-expert distribution
occurs in approximately $2.9\%$ of iterations.
During these spikes, Megatron-Kuai's MFU drops by $21.4\%$ on average
and AdaPipe's MFU drops by $18.2\%$ on average.
This is because they must fall back from under-provisioning strategy to an over-provisioning one, a process compounded by the cold-start overhead of OOM crashes.
\sysname{} transitions seamlessly between offloading and selective strategies
with no switching overhead,
maintaining stable MFU through spikes
and achieving an overall average speedup of $1.28\times$
over baselines during the spike phase. 

\noindent\textbf{Cross-Configuration Robustness.}
First, we evaluate configurations with different MoE sparsity.
We train models with the same total MoE parameter count
but different activated parameters,
which directly controls the per-GPU activation footprint.
Figure~\ref{fig:config-study}(a) shows the MFU of each system
across five configurations.
\sysname{} achieves the highest MFU in every configuration,
with an average speedup of $1.11\times$ over the best baseline. 
Notably, \sysname{}'s advantage grows with activation pressure:
with more activated parameters, more experts are activated per token,
which stresses baseline systems' static memory budgets
while \sysname{} adapts its reduction strategy to the increased pressure.
Additionally, more activated parameters imply greater compute density per iteration,
which naturally translates to higher absolute MFU
when the system can fully exploit the available memory.

Second, we evaluate configurations with varying input sequence length.
We use 423B-A30B and increase the input sequence length
from 4{,}096 to 16{,}384 tokens.
Longer sequences increase the per-token activation footprint
and the total memory required per micro batch.
Figure~\ref{fig:config-study}(b) shows the results.
All systems achieve their highest MFU at 4K tokens,
where activation memory is smallest relative to GPU capacity.
As sequence length increases to 8K and 16K,
MFU decreases across all systems because more aggressive activation reduction is required.
However, \sysname{}'s MFU degrades more gracefully:
its fine-grained mitigation avoids the all-or-nothing overhead
of baseline systems' coarse-grained strategies.


\subsection{Ablation Study}
\label{subsec:eval-ablation}

\begin{figure}
    \centering
    \includegraphics[width=\linewidth]{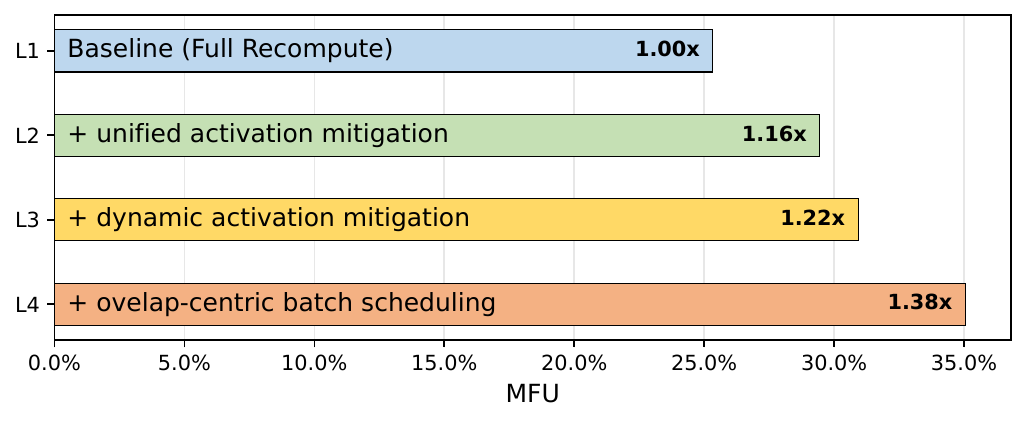}
    \vspace{-2em}
    \caption{Speedup breakdown from each optimization.} 
    \vspace{-2em}
    \label{fig:speedup-breakdown}
\end{figure}

\noindent\textbf{Speedup Breakdown.}
To quantify the contribution of each optimization,
We progressively enable \sysname{}'s components
starting from the baseline configuration.
Figure~\ref{fig:speedup-breakdown} shows the cumulative effect.
Each optimization contributes a positive, compounding speedup.
Static memory optimization (L1$\to$L2) provides a $1.16\times$ improvement
by enabling fine-grained, composable mitigation strategies
that replace coarse-grained layer-level policies.
Dynamic activation memory mitigation (L2$\to$L3) adds a further $1.05\times$
by dynamically adapting the strategy degree to per-iteration memory pressure.
Runtime overlap scheduling (L3$\to$L4) contributes an additional $1.13\times$
by hiding offload/reload and recomputation latency
through the dual-batch interleaving.
The combined effect of all three optimizations yields the full \sysname{} performance.





\noindent\textbf{Activation modeling accuracy.}
\sysname{}'s runtime decisions depend on the 
accurate modeling of the actual memory footprint.
We compare the predicted activation memory
against measured values across multiple scales and parallelism configurations.
\sysname{} achieves a mean prediction error of $1.56\%$
across all configurations, 
confirming that 
\S\ref{sec:model} faithfully depicts the memory behavior of MoE training activations.

\noindent\textbf{Runtime decision overhead.}
\sysname{} has three runtime components: dynamic memory budget assignment, activation mitigation strategy selection and DP-based batch scheduling.
We find memory budgets exhibit strong temporal locality ---
the pipeline phase repeats every $B_w$ micro-batches,
and MoE imbalance rates are stable for the majority of iterations ---
so most iterations hit the cache and skip algorithms entirely,
reducing the amortized overhead to near zero.
The total runtime overhead is less than $10$ ms per iteration,
while per iteration training time is more than $18$ seconds.


\section{Related Work}
\label{sec:related}

\noindent\textbf{MoE Model Training Systems.}
Early systems for MoE training focus on coarse-grained MoE architectures,
addressing computation and communication bottlenecks separately.
SmartMoE~\cite{smartmoe}
and FlexMoE~\cite{flexmoe_nie}
improve expert parallelism strategies to mitigate computation load imbalance.
Lina~\cite{lina}, FasterMoE~\cite{fastermoe} and Janus~\cite{liu2023janus}
optimize the all-to-all communication that dispatches tokens to experts,
while other work reduces communication overhead
through communication overlapping~\cite{overlap_comm_decomposition, coflow_scheduling, chen2024centauri, zhang2025comet, zheng2025tilelink}.
However, these systems all overlook the memory pressure
that arises during fine-grained MoE training,
where dynamic expert routing causes activation memory
to vary unpredictably across micro-batches.

\noindent\textbf{Memory-Efficient Training Systems.}
Reducing memory consumption during training has been studied extensively.
Early work targets single-GPU settings
through checkpointing~\cite{chen2016trainingdeepnetssublinear}
and offloading~\cite{swapadvisor, zong2023str, wen2022sdtrs}.
More recent systems address small-scale distributed training
by optimizing memory usage for parameters
and optimizer states~\cite{zhu2025mist,feng2023mobius}.
Another works specifically target activation memory,
but assumes static computation graphs
where activation sizes are known at compile time~\cite{megatron-kuai, adapipe, sun2025mepipe, li2021chimera, li2025hypertron, liu2023hanayo, liu2025mario, zong2025geodist, huang2025mtuner}.
None of these accounts for the dynamic activation memory
introduced by MoE routing,
where each micro-batch may produce a different activation footprint
depending on the runtime token-to-expert assignment.












\section{Conclusion}
\label{sec:conclusion}

We present \sysname{}, a large-scale MoE training system
that transforms activation memory management
from a static, one-time configuration choice
into continuous, per-batch control.
By combining fine-grained activation provisioning
with overlapped execution,
\sysname{} dynamically allocates exactly the activation memory
each training batch requires,
eliminating the over-provision vs.\ under-provision dilemma
that plagues existing systems.
\sysname{} scales to 12{,}288 GPUs on a model with 1040 billion parameters,
sustaining over 2~EFlops at 35\% MFU,
with average throughput gains of $1.14\times$ over Megatron-Kuai
and $1.13\times$ over AdaPipe.

\section{Acknowledgments}

We sincerely thank the anonymous reviewers for their valuable feedback on this paper. 
This work is supported by New Generation Artificial Intelligence-National Science and Technology Major Project under Grant 2025ZD0123801, NSFC for Distinguished Young Scholar under Grant 62225206, the National Natural Science Foundation of China (Grant No. 62402266, 62532006, and 62495062), and the Beijing Natural Science Foundation (Grant No. L2603007 and L242017).

\bibliographystyle{IEEEtran}
\bibliography{refs}

\end{document}